\documentstyle[preprint,aps]{revtex}

\begin{document}
\tightenlines
\draft

\title{A Beable Interpretation of the Continuous Spontaneous 
Localization Model}

\author{L. F. Santos $^1$ and C. O. Escobar $^2$}

\address{$^1$Departamento de F\'{\i}sica Nuclear \\
Instituto de F\'{\i}sica da Universidade de S\~ao Paulo, 
C.P. 66318, cep 05389-970 \\
S\~ao Paulo, S\~ao Paulo, Brazil\\
lsantos@charme.if.usp.br \\
$^2$ Departamento de  Raios C\'osmicos e Cronologia \\
Instituto de F\'{\i}sica Gleb Wataghin \\
Universidade Estadual de Campinas, C.P. 6165,  cep 13083-970\\
Campinas, S\~ao Paulo, Brazil\\
escobar@ifi.unicamp.br}

\maketitle

\begin{abstract}
We extend the beable interpretation, due to Bell, to the 
continuous spontaneous localization 
model (CSL). Results 
obtained by Vink are generalized to the modified Schr\"odinger equation
of Ghirardi, Pearle and Rimini (GPR), which allows 
a beable interpretation for both position 
and momentum. \\ 
\end{abstract}


{\bf 1. Introduction}

The usual interpretation of quantum mechanics deals fundamentally with results 
of measurements and therefore presupposes, besides a system, an apparatus 
to perform the measurements. However what the apparatus is and how to 
distinguish it from the system are questions with vague answers. In face of 
this problem, Bell \cite{Bell} proposed an interpretation in terms 
of `beables' 
instead of observables. Beables correspond to things that exist 
independently of the observation, therefore they can be 
assigned well defined values. In this 
way we avoid a cut between the microscopic 
(quantum) world and the macroscopic (classical) world.

Vink \cite{Vink} showed that two other well known interpretations of 
quantum mechanics - the causal interpretation associated with Bohm 
\cite{Bohm} and the stochastic interpretation due to Nelson \cite{Nelson} - 
are particular cases of the beable interpretation as developed by Bell. 
Moreover, he proposed that all observables, even those that do not 
commute, can attain beable status simultaneously.

In this paper we investigate the model proposed by GPR for the free 
particle \cite{GPR} from the beable point of view. 
We treat position and momentum as beables despite 
the fact that they are non-commuting and show that in the continuum 
limit (following Vink) they satisfy two stochastic differential 
equations, reproducing the results as obtained by 
Ghirardi, Rimini and Weber (GRW) for the average values 
and dispersions \cite{GRW}.

We start the paper by presenting Vink's description 
of causal and stochastic interpretations as particular cases of 
Bell's beables. After a brief exposition of the CSL model we obtain 
the main results of this work.

\vskip 1cm

{\bf 2. Causal and Stochastic Interpretations from the Beable 
Point of View}

In Bohm's causal interpretation the wave function $\psi $, which is a 
solution of the Schr\"odinger equation,  is a field that guides the particle, 
whose trajectory is obtained by solving the equation

\begin{equation}
\dot{x} = \frac{\nabla S(x,t)}{M}, 
\end{equation}
where $S/\hbar $ is the phase of the wave function written in the polar form

\begin{equation}
\psi (x,t) = R(x,t) \exp\left[ \frac{\imath S(x,t)}{\hbar }\right] .
\end{equation}
and $M$ is the particle mass.

In Nelson's stochastic approach particles 
play a preponderant role and are subjected 
to a stochastic process given by

\begin{equation}
d x =\left [ 2\nu \frac{\nabla R(x,t)}{R(x,t)} + 
\frac{\nabla S(x,t)}{M}\right] dt 
+ (2\nu )^{\frac{1}{2}} dw(t), 
\end{equation}
where $w(t)$ is a white noise: $<dw(t)> = 0$ and 
$<dw(t)^{2}> = dt$ with $\nu $ the diffusion constant given by 
$\nu = \hbar / 2M $. Note that when $\nu = 0$ 
we recover the causal interpretation.

Both approaches deal with trajectories, the particles have a 
definite position even if not observed, making position a beable 
in Bell's sense.

Vink \cite{Vink} showed a connection between the two approaches 
above and Bell's 
beable interpretation. Unlike Bell's approach, which is done 
in terms of fermion number, a discrete quantity, Vink shows that 
the beable concept can be extended to any observable in case it takes discrete 
values on small scales.

In case we want to find the trajectories of a set of commuting dynamical 
variables $O^i $, each one with $m$ discrete eigenvalues, we 
write the continuity equation in the $O$ representation as 

\begin{equation}
\partial _{t} P_{m} = \sum_{n} J_{mn},
\end{equation}
where the probability density $P_{m}$ and the source matrix $J_{mn}$ 
are defined by

\begin{equation}
P_{m} (t)  = |<O_{m} |\psi (t) >|^2 ,
\end{equation}

\begin{equation}
J_{mn} (t) = 2 Im \{<\psi (t)| O_{m} ><O_{m} |H| O_{n} > <O_{n} | 
\psi (t)> \}.
\end{equation}

We are using GPR notation $H = p^2 /2m\hbar + V(x)/\hbar $.

Following Bell, the  probability distribution 
of $O_{m} $ values, $P_{m} (t)$, satisfies the master equation

\begin{equation}
\partial _{t} P_{m} = \sum_{n} ( T_{mn} P_{n} - T_{nm} P_{m} ),
\end{equation}
where $T_{mn} dt$ is the transition probability expressing 
mathematically the probability for jumps from state $n$ to state $m$. 
To reconcile the quantum and stochastic views we equate (4) and (7):

\begin{equation}
J_{mn} = ( T_{mn} P_{n} - T_{nm} P_{m} ),
\end{equation}
with $T_{mn} \geq 0$ and  $J_{mn} = - J_{nm} $.

There is great freedom to find solutions of 
equation (8). Bell chooses a particular one 
for $n \neq m$,

\[ T_{mn} = \left\{
 \begin{array}{ll}
  J_{mn} /P_{n} &    J_{mn} > 0 \\
  0 &  J_{mn} \leq 0
 \end{array}
\right. \]

Restricting the position of a particle in one 
dimension to the sites of a lattice, $x=an$, with $n = 1,...,N$ and 
$a$ the lattice distance, it follows from the discrete version 
of the Schr\"odinger equation that $J_{mn} $ is given by

\begin{equation}
J_{mn} = \frac{1}{Ma } \{[S (an )]' P_{n} \delta _{n, m-1 } 
- [S (an)]' P_{n}  \delta _{n, m+1 }\} ,
\end{equation}
where use was made of the polar form of the wave function and 
$\psi (x+a) $ is expanded up to first order in $a$. In the expression 
above $[S(an)]' = [S(an+a) - S(an)]/a $.

For forward movement, Bell's choice becomes

\begin{equation}
T_{mn} = \frac{[S(an)]'}{Ma} \delta _{n, m-1}
\end{equation}
which gives the 
average displacement $dx = S(x)' dt/M$. In  the continuum limit, as
$a\rightarrow 0$, the particle has a velocity given by

\begin{equation}
\dot{x} = \frac{\nabla S(x)}{M}
\end{equation}
which corresponds to the result of the causal approach.

However, we could also add to $T_{mn} $ the solution of the homogeneous  
equation, $T^{o}_{mn} $

\begin{equation} 
T^{o}_{mn} P_{n} - T^{o}_{nm} P_{m} = 0
\end{equation}
for which we can choose a Gaussian with width $\sigma $

\begin{equation}
T^{o}_{mn} \propto
\exp \left\{ -\left[ m-n - \frac{2\sigma ln(P_{m} /P_{n} )}{4(m-n)} 
\right] ^2 /2\sigma \right\} .
\end{equation}

Assuming $\sigma $ sufficiently small, we can approximate 
$[\ln (P_{m} /P_{n})]/(m-n)$ by $2a[R(an)]'/R(an) $ arriving at the following 
Langevin equation for the particle position in the continuum limit

\begin{equation}
dx= \left[ (\beta 
\sigma a^2 ) \frac{\nabla R(x)}{R(x)} + \frac{\nabla S(x)}{ M} \right] dt + 
(\beta \sigma a^2 )^{\frac{1}{2} } dw,
\end{equation}
where $\beta $ is a free parameter

This equation coincides with Nelson's stochastic equation with 
$\beta \sigma a^2 = 2\nu $.

\vskip 1cm

{\bf 3. CSL Model}

In this model the wave function is subjected to a 
stochastic process in Hilbert space. In one dimension 
the evolution equation in the Stratonovich form is \cite{tail}

\begin{equation}
d\psi (x,t) =\left\{ [ -iH -\lambda ]dt  + \int dz dB(z,t) 
G(x-z) \right\} \psi (x,t)
\end{equation}
which does not preserve the norm of the wave function. A norm preserving 
evolution is described by the following Stratonovich equation \cite{Gisin,GPR}

\begin{equation}
d\phi (x,t) = \left\{ \left[-iH - \gamma \int dz  K(x,z)
 \right] dt + \left[\int dz dB(z,t) L(x,z)\right]\right\}  \phi (x,t).
\end{equation}
where

\begin{equation}
K(x,z) = (G(x-z))^2 [1 - 3 ||\psi ||^2 + 2 ||\psi ||^4 ]
\end{equation}
and

\begin{equation}
L(x,z) =  G(x-z)[1 - ||\psi ||^2 ]
\end{equation}

In the equations above $dB$ is  
a white noise ($<dB> = 0$ and $ <dB^2  >=  
\gamma dt$ ) and

\begin{equation}
G(x-z) = \sqrt{\frac{\alpha }{2\pi }} \exp \left[ -\alpha 
\frac{(x-z)^2]}{2}\right] 
\end{equation}
is an indication of the localization of the wave function.
The length parameter $\alpha $  and the frequency parameter 
$\lambda $ are related to $\gamma $ according to $\gamma =
\lambda (4\pi /\alpha )^{\frac{1}{2} }$.
They are chosen in such a way that the 
new evolution equations do not give different results from the usual 
Schr\"odinger unitary evolution for microscopic systems with few degrees 
of freedom, but when a macroscopic system is described 
there is a fast decay of the 
macroscopic linear superpositions which are quickly transformed 
into statistical mixtures \cite{GPR,GRW}.

We now formulate this model in terms of beables.
The reader may be worried that Vink's 
approach does not hold for a non-linear evolution 
such as in equation (16), however the 
non-linear terms only involve expectation values in 
the given state and this equation gives the increment 
$d\phi $ for a given $dt$. All we have to do in order 
to adapt Vink's approach to our case is to compute (4) 
in the following way: $dP_{m} / dt = \sum J_{nm} $.
Taken this precaution into account
a solution of equation (8) for the CSL model can be given as

\begin{equation}
T^{x}_{mn} dt =  \frac{J^{x}_{mn} }{P_{n} } dt + \beta  T^{ox}_{mn} dt
\end{equation}
where, for the modified Schr\"odinger equation (16) 
\footnote{Notice that as the beable approach deals with 
transition probabilities we use the norm preserving evolution equation 
(16).} and forward movement, 
$J^{x}_{mn} $ \footnote{The superscript $x$ is used to remind 
that position is the beable.} is given by

\begin{equation}
J^{x}_{mn} =  \frac{1}{Ma } [S (an )]' 
P_{n} \delta _{n, m-1 } +2\left( -\gamma \int dz K(x,z)
  + \int dz \frac{\partial B(z,t)}{\partial t} L(x,z)\right) P_{n} 
\delta _{m,n}
\end{equation}
The last term in equation (21)
comes from the non-unitary part of the evolution of 
the wave function (16), but does not contribute to 
the displacement $dx$ and consequently $dx$ obeys an equation of 
the same form as equation (14). However, the wave function is now 
different from the one unitarily evolved by the action of $H$, 
$\psi _{S} (x,t)$. In order to obtain the normalized wave function 
$\phi (x,t)$ it is easier to solve equation (15) than the non-linear 
equation (16) and then use \cite{GPR,Ghirardi}: 
$\phi (x,t) = \psi (x,t)/||\psi || $, which gives

\begin{equation}
\psi (x,t) = \frac{\psi _{S} (x,t)}{||\psi ||} \exp \left[-\lambda t \right]
\exp \left\{ \int _{0}^{t} \left[ \int dz G(x-z) dB(z,t')\right] \right\} ,
\end{equation}

Notice that the norm $||\psi ||$ does not depend on $x$, therefore the 
stochastic differential equation for the free particle is

\begin{equation}
dx =\frac{p_{o}}{M} dt 
+ 2\nu  
\left\{ \int _{0}^{t} \left[ \int dz dB(z,t') 
\frac{\partial G(x-z)}{\partial x}\right]\right\} dt   + 
(2\nu )^{\frac{1}{2}} dw.
\end{equation}
where $dw$ and $dB$ are two independent white noises.

In equation (23) the first term on the right hand side 
describes a single free particle
deterministic evolution. The two  
other terms describe the stochastic processes, with the last one being a 
standard diffusion and the second term, a non standard diffusion 
which exhibits the non-locality of the localization process. This 
second term indicates that the particle position tracks 
the wave function. The position increment induced by this term drives 
the particle to where the wave function is increasing according to 
the fluctuating term in equation (22).

Considering now momentum as a beable, we 
repeat the procedure for position and restrict it to the sites 
of a lattice, $p=bn$, with $n = 1, ..., N$ and $b$ the lattice 
distance. The norm preserving evolution equation in the $p$ representation

\begin{eqnarray}
\frac{\partial \tilde{\phi }(p,t)}{\partial t} &=& -iH (p) 
\tilde{\phi }(p,t)  -
\sqrt{\frac{1}{2 \pi  \hbar }} 
\int dp' dx dz  \gamma  
K(x,z)
\exp \left[- i\frac{(p-p')x}{\hbar } 
\right] \tilde{\phi }(p',t) + \\ \nonumber
                                       &+&
\sqrt{\frac{1}{2 \pi  \hbar }} 
\int dp' dx dz  \frac{\partial B(z,t)}{\partial t} L(x,z)
\exp \left[- i\frac{(p-p')x}{\hbar } 
\right] \tilde{\phi }(p',t)
\end{eqnarray}
after being discretized gives

\begin{eqnarray}
J^{p}_{mn} &=& \sqrt{\frac{1}{2 \pi  \hbar }} 
 \sum _{m'} \int dx dz 
\left[- \gamma K(x,z) + \frac{\partial B(z,t)}{\partial t} L(x,z) \right]
 \\ \nonumber
           & & \left\{ e^{\left[ -ib\frac{(m-m')x}{\hbar }
\right]} <\tilde{\phi }(p,t) |O_{m}>  <O_{n}|O_{m'}> 
+ e^{\left[ ib \frac{(m-m')x}{\hbar } \right]} 
<O_{m'} |O_{n} > <O_{m} |\tilde{\phi }(p,t)>\right\} \delta _{m,n}
\end{eqnarray}

The transition probability for jumps in momentum is

\begin{equation}
T^{p}_{mn} dt  = \frac{J^{p}_{mn} }{P_{n} } dt + \xi T^{op}_{mn} dt
\end{equation}
where

\begin{equation}
T^{op}_{mn} \propto
\exp \left\{ -\left[ m-n - \frac{2\Omega ln(P_{m} /P_{n} )}{4(m-n)} 
\right] ^2 /2\Omega \right\}
\end{equation}

Assuming $\Omega $ small, we can approximate  
$[\ln (P_{m} /P_{n})]/(m-n)$ by \\ 
$2b[\tilde{\phi }^{*}(bn) \tilde{\phi }(bn)]'/
[\tilde{\phi }^{*} (bn) \tilde{\phi }(bn)]$ and obtain for $dp$

\begin{equation}
dp = \left[ (\xi \Omega b^2 )\frac{\nabla _{p}
 [\tilde{\phi }^{*} (p) \tilde{\phi }(p)]}
{\tilde{\phi }^{*} (p) \tilde{\phi }(p) }\right] dt +  
(\xi \Omega b^2 )^{\frac{1}{2}}dw
\end{equation}

The localization process in configuration space leads to a spreading 
of the wave-packet in momentum space, which in turn, removes the 
momentum dependence on $|\tilde{\phi (p)}|^{2} $, eliminating 
the first term in equation (27). 

In order to obtain the same mean values and dispersions for position and 
momentum in the GRW model for a free particle, we need to use
$\xi \Omega b^2 = \hbar ^{2} \alpha \lambda /2$ and can thus write

\begin{equation}
dp = \hbar \sqrt{\frac{\alpha \lambda }{2}} dw
\end{equation}

Notice that the stochastic process for momentum is a consequence of the 
collapse of the wave function, which vanishes when GRW parameters 
($\alpha $, $\lambda $) go to zero.

The two stochastic differential equations for position and momentum 
(eqs. 23 and 29) lead to the same Fokker-Planck equation for the 
phase-space density as obtained by GRW.

\vskip 1 cm 

{\bf 4. Concluding Remarks}

We have given a beable interpretation for the CSL model which leads
quite naturally to simultaneous beable status for both position and 
momentum.

The localization of the wave function, in addition to inducing fluctuation 
in momentum (eq. 29), introduces a new stochastic process for the 
displacement (eq. 23), which drives the particle to where the wave function 
localizes.

We have treated in this paper the free particle case and intend to extend 
this treatment to other cases in a future publication.

\acknowledgments
The authors acknowledge the support of the Brazilian 
Research Council, CNPq.

\end{document}